\documentstyle[preprint,eqsecnum,aps]{revtex}

\begin{document}
\draft
\preprint{TD Bank and UW}
\title{Hedging The Risk In The Continuous Time \\Option Pricing Model
With Stochastic Stock Volatility 
}
\author{D. F. Wang$^{1,2}$}
\address{$^1$Department of Statistics and Actuarial Science\\
University of Waterloo, 
Waterloo, ONT N2L 3E5 Canada\\
$^2$Toronto Dominion Bank}
\date{June, 1998}
\maketitle
\begin{abstract}

In this work, I address the issue of forming riskless hedge in
the continuous time option pricing model with stochastic stock volatility.
I show that it is essential to verify whether the replicating
portfolio is self-financing, in order for the theory to be
self-consistent. The replicating methods in existing finance
literature are shown to violate the self-financing constraint
when the underlying asset has stochastic volatility. Correct
self-financing hedge is formed in this article. 
\end{abstract}
\pacs{PACS number: }


It has been indicated by empirical observations that stock price
volatility is stochastic. Considerable amount of analytical and
numerical work has been devoted to pricing derivatives when
the underlying stock has 
stochastic volatility\cite{merton1,garman,scott,hull1,hull2,wigg,boyle,book,heston,ba,ss,ball}. 
For example, there were early works done by Merton\cite{merton1}, and that by Garman etc\cite{garman}.
In current existing finance literature, various authors constructed
riskless hedging portfolios in different ways\cite{scott,ba}. With the principle of
no-arbitrage, the riskless hedging portfolio should match the return
of a riskless loan. This will result in the partial differential equation
satisfied by the option prices. In the situation where the underlying
asset has stochastic volatility, investors' preferences, such as the risk
preminum on the stock, will get involved explicitly. 

In this work, I would like to address the issue of hedging the risk
for the above system. In the continuous time models, it is 
necessary to make sure that the replicating portfolio is self-financing
as it was assumed to be. This verification is essential for the theory
to be self-consistent. In the following, I show that in spite of the final
correctness of the derived PDE for option prices, the hedging strategies
in some current finance literature\cite{scott,ba}
turned out to violate the self-financing condition that was assumed
to hold. Arguments and corrections are given in this work when the underlying
asset has stochastic volatility.

Below, we follow the reference\cite{scott} and its notations. 
Let us first define 
a probability space $(\Omega, Q, F)$. Consider 
the stock price obeying the stochastic process
\begin{equation}
dP=\alpha P dt +\sigma P dZ_1,
\end{equation} 
where the volatility $\sigma$ is described by another mean-reverting
process
\begin{equation}
d\sigma=\beta (\bar \sigma -\sigma)dt +\gamma dZ_2. 
\end{equation} 
Here, both $Z1$ and $Z_2$ are one dimensional Brownian motions. 
The co-quadratic process of $Z_1$ and $Z_2$ is assumed to be
$[Z_1,Z_2] =t\delta $. For the process $\sigma$ described above, there is
non-zero chance for the volatility to be negative. However,
the following argument of ours will remain unchanged when other 
positive processes for the volatility are used, such as for the process
$dln\sigma = \beta (\bar {ln\sigma}-ln\sigma)dt +\gamma dZ_2$.

To hedge away the risk, a portfolio was constructed to have two call options 
and one stock\cite{scott}. The two options have different maturity dates $T_1$ and $T_2$.
Denote $\tau_1=T_1-t$ and $\tau_2=T_2-t$. The option price was assumed to be 
$H(P, \sigma, \tau)$, 
a function of stock price $P$, the stock volatility $\sigma$, and $\tau=T-t$.
In the following, we may use short notations 
$H(\tau_1)=H(P, \sigma, \tau_1), H(\tau_2)=H(P, \sigma, \tau_2)$.  
According to the reference\cite{scott}, a trading strategy was proposed to be 
\begin{equation}
\phi=(\phi_1, \phi_2, \phi_3)=(1, \omega_2, \omega_3) 
\end{equation}
where one has 
\begin{eqnarray}
&&\phi_1=1,\nonumber\\
&&\phi_2=\omega_2=-{H_2(P,\sigma,\tau_1)\over H_2(P,\sigma,\tau_2)},\nonumber\\
&&\phi_3=\omega_3=-H_1(P,\sigma,\tau_1)+
{H_2(P,\sigma,\tau_1) H_1(P,\sigma,\tau_2)\over H_2(P,\sigma,\tau_2)}.
\label{eq:trading} 
\end{eqnarray} 
Here one uses the conventions $H_2(P,\sigma,\tau)=\partial H(P,\sigma,\tau)/\partial\sigma,
H_1(P,\sigma,\tau)=\partial H(P,\sigma,\tau)/\partial P, H_3(P,\sigma,\tau)=
\partial H(P,\sigma,\tau)/\partial \tau=
- \partial H(P,\sigma,\tau)/\partial t$. 
The wealth process of this portfolio is thus given by 
$V(\phi)=\phi_1 H(\tau_1) + \phi_2 H(\tau) +\phi_3 P$. 
Assuming this portfolio is self-financing, the change of the portfolio value will
therefore take the following form:
\begin{equation}
dV=\phi_1 dH(\tau_1) +\phi_2 dH(\tau_2) +\phi_3 dP, 
\end{equation} 
which can be seen being riskless after substituting the trading strategy explicitly.
This change of the wealth process should match the return of a riskless loan, leading
to a partial differential equation satisfied by the option price. Combining this
with some general equilibrium consideration, one therefore obtains the option valuation
PDE when the underlying stock volatility is stochastic\cite{scott}.

However, we would like to note that it is essential to fully justify the above
assumption that the corresponding portfolio is indeed self-financing. In spite of the
fact that the final PDE based on this portfolio is correct accidentally, it is essential
to check explicitly $dV=\phi_1 dH(\tau_1) +\phi_2 dH(\tau_2) +\phi_3 dP$ holds 
within the continuous-time framework for the theory to be self-consistent. 
It is shown below that this particular trading strategy defined by Eq.(~\ref{eq:trading})
is not self-financing.   

Suppose that we have found the price of European call option from Eq.(4) of the reference\cite{scott},
with appropriate boundary condition $H(P,\sigma, \tau)=max(0,P(T)-K)$ at time $t=T$. 
Let us substitute this back into the trading strategy $\phi$ defined before. The wealth
is $V=V(\phi)=\phi_1 H(\tau_1) + \phi_2 H(\tau) +\phi_3 P$. Therefore, from simple rule of 
stochastic calculus, we should have
\begin{equation}
dV=dH(\tau_1)+\omega_2 dH(\tau_2)+\omega_3 dP+ ( H(\tau_2)d\omega_2+d[\omega_2,H(\tau_2)]
+P d\omega_3 +d[P,\omega_3]). 
\end{equation} 
Denote $dW= H(\tau_2)d\omega_2+d[\omega_2,H(\tau_2)]+ P d\omega_3 +d[P,\omega_3]$. 
If we find that $dW=0$, then, the original trading strategy is self-financing.
Otherwise, the proposed trading strategy is not self-financing, and there is 
self-inconsistence in the theory. We can compute $dW$ explicitly, although the computation
is a bit tedious. In general, $dW$ will take the form
\begin{equation}
dW=f^{(0)}(P,\sigma,\tau_1,\tau_2) dt +f^{(1)}(P,\sigma,\tau_1,\tau_2)dZ_1 + 
f^{(2)}(P,\sigma,\tau_1,\tau_2) dZ_2,
\end{equation}
where the functions $f^{(0)}, f^{(1)}, f^{(2)}$ can be found, given the European
option price. 
One can easily see that $d[\omega_2,H(\tau_2)]= K_1(P,\sigma,\tau_1,\tau_2) dt$ 
and $d[P,\omega_3]=K_2(P,\sigma,\tau_1,\tau_2) dt$, where both $K_1$ and $K_2$ are
some functions of $P, \sigma, \tau_1,$ and $\tau_2$. Hence, $ f^{(1)}$ and $ f^{(2)}$
solely come from the contributions of expanding $H_2(\tau_2) d\omega_2 $ and $P d\omega_3$.
Expanding these two terms will also contribute to $f^{(0)}$.
After some tedious calculation, we obtain $f^{(1)}dZ_1+f^{(2)}dZ_2$ explicitly.
The first coefficient is found to be: 
\begin{eqnarray}
f^{(1)}=&&(H (\tau_2)-PH_1(\tau_2))(-{H_{21}(\tau_1)\over H_2(\tau_2)} + 
{H_2(\tau_1)\over H_2^2(\tau_2)} H_{21}(\tau_2))\sigma P +\nonumber\\ 
&&-PH_{11}(\tau_1)\sigma P +P{H_2(\tau_1)
\over H_2(\tau_2)} H_{11}(\tau_2) \sigma P. 
\end{eqnarray}
We have also found the second coefficient:
\begin{eqnarray}
f^{(2)}=(H(\tau_2)-PH_1(\tau_2)) (-{H_{22}(\tau_1)\over H_2(\tau_2)} +{H_2(\tau_1)\over
H_2^2(\tau_2)} H_{22}(\tau_2)) \gamma + \nonumber\\
-PH_{12}(\tau_1)\gamma + P {H_2(\tau_1)\over H_2(\tau_2)} H_{12}(\tau_2) \gamma . 
\end{eqnarray}
When $\tau_1=\tau_2$, we see that $f^{(1)}=f^{(2)}=0$. However, for general
$\tau_1\ne \tau_2$, it is true that 
\begin{equation}
f^{(1)}\ne 0 , f^{(2)} \ne 0,
\end{equation}
indicating that in general $dW$ is non-zero (almost surely) in the probability
space $(\Omega,Q, F)$. One observes that the hedging portfolio constructed by 
reference\cite{scott} is not self-financing in the continuous time framework. 

In order to remedy the situation, we may construct a trading strategy in different way:
\begin{equation}
\phi=(\phi_1,\phi_2,\phi_3)=({Vx_1\over H(\tau_1)}, {Vx_2\over H(\tau_2)}, {Vx_3\over P})
\end{equation} 
where $x_1+x_2+x_3=1$ is assumed, and $V=V(P,\sigma,\tau_1,\tau_2)$ is to be determined.
We choose the x's to be 
\begin{equation}
{x_1\over H(\tau_1)}=a, {x_2\over H(\tau_2)}=a \omega_2  , {x_3\over P}=\omega_3 a, 
\end{equation}
with $a=1/[ H(\tau_1) +\omega_2 H(\tau_2) + P \omega_3]$, and with $\omega_2, 
\omega_3$ defined as before. The process $V(P,\sigma,\tau_1,\tau_2)$ is defined 
by the differential equation
\begin{eqnarray}
{dV\over V}=&&(\phi_1 H(\tau_1) + \phi_2 H(\tau_2) +\phi_3 P)^{-1} \times
[ -H_3(\tau_1) +{1\over 2} H_{11}(\tau_1)\sigma^2P^2 +H_{12}(\tau_1)\delta \gamma \sigma 
P +{1\over 2} H_{22}(\tau_1) \gamma^2 -\nonumber\\
&&-{H_2(\tau_1)\over H_2(\tau_2)}
[-H_3(\tau_2)+{1\over 2}H_{11}(\tau_2) \sigma^2 P^2 +H_{12}(\tau_2)\delta \gamma 
\sigma P +{1\over 2} H_{22}(\tau_2) \gamma^2]]dt. 
\end{eqnarray}
Suppose that we have obtained European call option price $H(P,\sigma,\tau)$ from
Eq.(4) of the reference\cite{scott} with proper boundary condition. Substitute $H$ into the above 
equation and it will determine the finite-variation process $V$. 
With this $V$ substituted into the trading strategy as defined above, 
one can easily check that the self-financing constraint is satisfied.
Therefore, there will be no inconsistency in the derivation of PDE for option
pricing.

In one recent interesting article, the path-integral technique was used to do perturbation
for the option pricing model when the underlying asset has stochastic volatility\cite{ba}. 
However, we would like to note that for constant volatility, the usual way in 
section one of the paper\cite{ba} to construct the riskless portfolio of one long option  $f$
and $\partial f/\partial S$ share short stock is not self-financing trading strategy, as 
shown by Bergman, and by Musiela $\&$ Rutkowski\cite{bergman,book}. 
Apart from this, when the volatility is 
random, it is known in finance community that in general a riskless hedge can not be formed from
only one option and the stock. The riskless hedge 
constructed with one option and
one stock in the paper\cite{ba} 
is always zero strategy. First, one can check that  
Eq.(8e) in the reference\cite{ba} would give $\theta_1=0$, because $\partial f/\partial S \ne 0$,
and $\partial f/\partial V\ne 0$ ( in his notations ). 
Therefore, the same equation again will give $\theta_2=0$. Hence the trading strategy is
always zero. We wish to note that as Wiggins and other people did, one could only form
a portfolio of one stock and one option, such that the portfolio has zero co-quadratic
variation with the stock\cite{wigg}. It is impossible to form a riskless hedge with only
one option and one stock in this case in general.  

In summary, we have addressed the issue of forming a riskless hedge in continuous time
option pricing model when the underlying asset has stochastic volatility. 
It is essential to the self-consistency of theory to explicitly justify whether
the hedge is indeed self-financing as one assumed. Some confusion in existing
finance literature on this aspect is cleared.   

I would like to thank Professors P. Boyle and D. McLeish for 
the finance theories I learned from them. Conversations with
Prof. K. S. Tan of UW, 
Dr. Hou-Ben Huang and Dr. Z. Jiang of TD Securities, Dr. Bart Sisk of TD Bank,
Dr. Daiwai Li
and Dr. Craig Liu of Royal Bank, Dr. ChongHui Liu  and Dr. J. Faridani 
of Scotia Bank, are gratefully knowledged.  
Any errors of this article are mine. 

Email address: d6wang@barrow.uwaterloo.ca. 
This paper
has been submitted to Journal of Financial and Quantatitive Analysis for publication.

\end{document}